\documentclass[aps,prl,english,twocolumn,superscriptaddress,showpacs,floats]{revtex4}
\usepackage{epsfig,epsf,amsmath,amsthm,amsfonts,
            graphics,graphicx,amssymb,babel}

\input psfig.sty

\def\met{\mbox{${\hbox{$E$\kern-0.6em\lower-.1ex\hbox{/}}}_T$}} 

\newcommand{\Tesla}{\ensuremath{\mathrm{T}}}
\newcommand{\TeV}{\ensuremath{\mathrm{Te\kern -0.1em V}}}
\newcommand{\GeV}{\ensuremath{\mathrm{Ge\kern -0.1em V}}}
\newcommand{\MeV}{\ensuremath{\mathrm{Me\kern -0.1em V}}}
\newcommand{\GeVCSq}{\ensuremath{\GeV\!/c^2}}
\newcommand{\GeVC}{\ensuremath{\GeV\!/c}}
\newcommand{\et}{E_T}
\newcommand{\pt}{p_T}
\newcommand{\ppbar} {p\bar{p}}
\newcommand{\ttbar} {t\bar{t}}

\newcommand{\fbinv} {fb$^{-1}$}
\newcommand {\abseta} {|\eta|}

\newcommand{\ntwo}{\tilde{{\chi}}^{0}_{2}}
\newcommand{\cone}{\tilde{{\chi}}^\pm_{1}}
\newcommand{\none}{\tilde{{\chi}}^{0}_{1}}

\begin{document}
\lefthyphenmin=2
\righthyphenmin=3

%
%
\title{
Search for Chargino-Neutralino Production in $\ppbar$ Collisions at
$\sqrt{s}=1.96$~$\TeV$
}

\affiliation{Institute of Physics, Academia Sinica, Taipei, Taiwan 11529, Republic of China} 
\affiliation{Argonne National Laboratory, Argonne, Illinois 60439} 
\affiliation{Institut de Fisica d'Altes Energies, Universitat Autonoma de Barcelona, E-08193, Bellaterra (Barcelona), Spain} 
\affiliation{Baylor University, Waco, Texas  76798} 
\affiliation{Istituto Nazionale di Fisica Nucleare, University of Bologna, I-40127 Bologna, Italy} 
\affiliation{Brandeis University, Waltham, Massachusetts 02254} 
\affiliation{University of California, Davis, Davis, California  95616} 
\affiliation{University of California, Los Angeles, Los Angeles, California  90024} 
\affiliation{University of California, San Diego, La Jolla, California  92093} 
\affiliation{University of California, Santa Barbara, Santa Barbara, California 93106} 
\affiliation{Instituto de Fisica de Cantabria, CSIC-University of Cantabria, 39005 Santander, Spain} 
\affiliation{Carnegie Mellon University, Pittsburgh, PA  15213} 
\affiliation{Enrico Fermi Institute, University of Chicago, Chicago, Illinois 60637} 
\affiliation{Comenius University, 842 48 Bratislava, Slovakia; Institute of Experimental Physics, 040 01 Kosice, Slovakia} 
\affiliation{Joint Institute for Nuclear Research, RU-141980 Dubna, Russia} 
\affiliation{Duke University, Durham, North Carolina  27708} 
\affiliation{Fermi National Accelerator Laboratory, Batavia, Illinois 60510} 
\affiliation{University of Florida, Gainesville, Florida  32611} 
\affiliation{Laboratori Nazionali di Frascati, Istituto Nazionale di Fisica Nucleare, I-00044 Frascati, Italy} 
\affiliation{University of Geneva, CH-1211 Geneva 4, Switzerland} 
\affiliation{Glasgow University, Glasgow G12 8QQ, United Kingdom} 
\affiliation{Harvard University, Cambridge, Massachusetts 02138} 
\affiliation{Division of High Energy Physics, Department of Physics, University of Helsinki and Helsinki Institute of Physics, FIN-00014, Helsinki, Finland} 
\affiliation{University of Illinois, Urbana, Illinois 61801} 
\affiliation{The Johns Hopkins University, Baltimore, Maryland 21218} 
\affiliation{Institut f\"{u}r Experimentelle Kernphysik, Universit\"{a}t Karlsruhe, 76128 Karlsruhe, Germany} 
\affiliation{High Energy Accelerator Research Organization (KEK), Tsukuba, Ibaraki 305, Japan} 
\affiliation{Center for High Energy Physics: Kyungpook National University, Taegu 702-701, Korea; Seoul National University, Seoul 151-742, Korea; SungKyunKwan University, Suwon 440-746, Korea} 
\affiliation{Ernest Orlando Lawrence Berkeley National Laboratory, Berkeley, California 94720} 
\affiliation{University of Liverpool, Liverpool L69 7ZE, United Kingdom} 
\affiliation{University College London, London WC1E 6BT, United Kingdom} 
\affiliation{Centro de Investigaciones Energeticas Medioambientales y Tecnologicas, E-28040 Madrid, Spain} 
\affiliation{Massachusetts Institute of Technology, Cambridge, Massachusetts  02139} 
\affiliation{Institute of Particle Physics: McGill University, Montr\'{e}al, Canada H3A~2T8; and University of Toronto, Toronto, Canada M5S~1A7} 
\affiliation{University of Michigan, Ann Arbor, Michigan 48109} 
\affiliation{Michigan State University, East Lansing, Michigan  48824} 
\affiliation{University of New Mexico, Albuquerque, New Mexico 87131} 
\affiliation{Northwestern University, Evanston, Illinois  60208} 
\affiliation{The Ohio State University, Columbus, Ohio  43210} 
\affiliation{Okayama University, Okayama 700-8530, Japan} 
\affiliation{Osaka City University, Osaka 588, Japan} 
\affiliation{University of Oxford, Oxford OX1 3RH, United Kingdom} 
\affiliation{University of Padova, Istituto Nazionale di Fisica Nucleare, Sezione di Padova-Trento, I-35131 Padova, Italy} 
\affiliation{LPNHE, Universite Pierre et Marie Curie/IN2P3-CNRS, UMR7585, Paris, F-75252 France} 
\affiliation{University of Pennsylvania, Philadelphia, Pennsylvania 19104} 
\affiliation{Istituto Nazionale di Fisica Nucleare Pisa, Universities of Pisa, Siena and Scuola Normale Superiore, I-56127 Pisa, Italy} 
\affiliation{University of Pittsburgh, Pittsburgh, Pennsylvania 15260} 
\affiliation{Purdue University, West Lafayette, Indiana 47907} 
\affiliation{University of Rochester, Rochester, New York 14627} 
\affiliation{The Rockefeller University, New York, New York 10021} 
\affiliation{Istituto Nazionale di Fisica Nucleare, Sezione di Roma 1, University of Rome ``La Sapienza," I-00185 Roma, Italy} 
\affiliation{Rutgers University, Piscataway, New Jersey 08855} 
\affiliation{Texas A\&M University, College Station, Texas 77843} 
\affiliation{Istituto Nazionale di Fisica Nucleare, University of Trieste/\ Udine, Italy} 
\affiliation{University of Tsukuba, Tsukuba, Ibaraki 305, Japan} 
\affiliation{Tufts University, Medford, Massachusetts 02155} 
\affiliation{Waseda University, Tokyo 169, Japan} 
\affiliation{Wayne State University, Detroit, Michigan  48201} 
\affiliation{University of Wisconsin, Madison, Wisconsin 53706} 
\affiliation{Yale University, New Haven, Connecticut 06520} 
\author{T.~Aaltonen}
\affiliation{Division of High Energy Physics, Department of Physics, University of Helsinki and Helsinki Institute of Physics, FIN-00014, Helsinki, Finland}
\author{A.~Abulencia}
\affiliation{University of Illinois, Urbana, Illinois 61801}
\author{J.~Adelman}
\affiliation{Enrico Fermi Institute, University of Chicago, Chicago, Illinois 60637}
\author{T.~Affolder}
\affiliation{University of California, Santa Barbara, Santa Barbara, California 93106}
\author{T.~Akimoto}
\affiliation{University of Tsukuba, Tsukuba, Ibaraki 305, Japan}
\author{M.G.~Albrow}
\affiliation{Fermi National Accelerator Laboratory, Batavia, Illinois 60510}
\author{S.~Amerio}
\affiliation{University of Padova, Istituto Nazionale di Fisica Nucleare, Sezione di Padova-Trento, I-35131 Padova, Italy}
\author{D.~Amidei}
\affiliation{University of Michigan, Ann Arbor, Michigan 48109}
\author{A.~Anastassov}
\affiliation{Rutgers University, Piscataway, New Jersey 08855}
\author{K.~Anikeev}
\affiliation{Fermi National Accelerator Laboratory, Batavia, Illinois 60510}
% ...
\collaboration{CDF Collaboration\footnote{With visitors from $^a$University of Athens, 15784 Athens, Greece, 
$^b$University of Bristol, Bristol BS8 1TL, United Kingdom, 
$^c$University Libre de Bruxelles, B-1050 Brussels, Belgium, 
$^d$Cornell University, Ithaca, NY  14853, 
$^e$University of Cyprus, Nicosia CY-1678, Cyprus, 
$^f$University College Dublin, Dublin 4, Ireland, 
$^g$University of Edinburgh, Edinburgh EH9 3JZ, United Kingdom, 
$^h$University of Heidelberg, D-69120 Heidelberg, Germany, 
$^i$Universidad Iberoamericana, Mexico D.F., Mexico, 
$^j$University of Manchester, Manchester M13 9PL, England, 
$^k$Nagasaki Institute of Applied Science, Nagasaki, Japan, 
$^l$University de Oviedo, E-33007 Oviedo, Spain, 
$^m$University of London, Queen Mary College, London, E1 4NS, England, 
$^n$University of California Santa Cruz, Santa Cruz, CA  95064, 
$^o$Texas Tech University, Lubbock, TX  79409, 
$^p$University of California, Irvine, Irvine, CA  92697, 
$^q$IFIC(CSIC-Universitat de Valencia), 46071 Valencia, Spain, 
}}
\noaffiliation

%
%
\begin{abstract}
We present the results of a search for associated production of 
the chargino and neutralino supersymmetric particles using up 
to $1.1$~fb$^{-1}$ of integrated luminosity
collected by the CDF~II experiment at the Tevatron $\ppbar$ collider 
at a center-of-mass energy of 1.96~TeV. The search is conducted by analyzing 
events with 
a large transverse momentum imbalance and either 
three charged leptons or two charged leptons of 
the same electric charge.
The numbers of observed events are found to be consistent with 
standard model expectations. Upper limits on the production cross section
are derived in different theoretical models. In one of these models
a lower limit on the mass of the chargino is set at $129$~GeV/$c^2$ at
the 95\% confidence level.

\end{abstract}

\maketitle

Supersymmetry (SUSY)~\cite{susy} is one of the most appealing and most studied
theories for physics beyond the standard model (SM). SUSY predicts the 
existence of a super-partner particle (sparticle) for each SM particle, 
sharing the same quantum numbers but differing by half a unit of spin.
SUSY addresses several problems of the SM: it
can solve the ``hierarchy problem''~\cite{hierarchy} 
(allowing the electroweak force and the gravitational force 
to coexist naturally), it can provide a good candidate
for the cold dark matter in the universe~\cite{dm} and it makes
possible a unification of the fundamental
forces at high energies~\cite{unification}. Since no SUSY particles have been observed so far
their masses must be much higher than those of their SM partners, and thus
supersymmetry, if it exists, must be a broken symmetry. 

A natural solution to the hierarchy problem and the
prospect of gauge coupling unification 
suggest that sparticle masses are near the 
electroweak scale and thus may be observable at the Tevatron $\ppbar$ 
collider.
We present a search for the associated production of
the lightest chargino $\cone$ and the
second-to-lightest neutralino $\ntwo$, which are 
mass eigenstates of the super-partners of the electroweak gauge and Higgs 
bosons.

Chargino-neutralino production, $\ppbar \to \cone \ntwo +X$, 
is one of the most interesting SUSY processes at the Tevatron 
in models in which the gauginos are light.
This process can be detected
through the observation of isolated charged leptons~\cite{cdfcoord} and a 
large imbalance in the transverse energy (\met)
from the decays $\cone \to \ell^\pm \nu \none$ and
$\ntwo \to \ell^+ \ell^- \none$, where $\ell = e,\mu,\tau$
and $\none$ is the lightest SUSY particle, 
which is assumed to be stable and to escape detection.
This signature has the experimental advantage that at hadron colliders
leptons are relatively rare compared to the copiously produced jets, they
are well identifiable, and the SM backgrounds are rather small as 
they arise primarily from electroweak processes. 
We use three benchmark models based on the minimal supersymmetric standard 
model (MSSM) to interpret the data. The models differ mostly 
in the leptonic branching ratios of the $\cone$ and $\ntwo$ and in the 
kinematic properties of the leptons. 

Previous searches at LEP excluded chargino masses below 
$103.5$~GeV/$c^2$~\cite{lepcone} and masses of the lightest 
neutralino $\none$ below 50.3~\GeVCSq~\cite{lepnone} at the
95\% confidence level (CL). These constraints are very robust and do not change much  
within minimal supergravity-inspired SUSY models.
The D0 collaboration recently constrained the 
lightest chargino mass to be larger than 117~GeV/$c^2$ at 95\%~CL 
in a specific scenario~\cite{d0trilep}.

%
This letter summarizes and interprets the results from seven
individual search channels with different lepton flavors and 
kinematic properties in the final state. The analyses use 
$\ppbar$ collision data taken by the CDF~II
detector at the Tevatron accelerator with a center-of-mass
energy of $\sqrt{s}=1.96$~TeV. The data, collected between March 
2002 and February 2006, correspond to an integrated luminosity 
between $0.7$ and  $1.1$~\fbinv, depending on the decay signature.

The CDF~II detector~\cite{cdf} is 
cylindrically symmetric around the beam-pipe in which the 
protons and anti-protons 
collide~\cite{cdfcoord}. The transverse momentum of charged particles, $\pt$, 
is measured by a 
tracking system composed of an eight-layer silicon
strip detector~\cite{silicon} and a 96-layer drift chamber~\cite{COT};
both are located inside a solenoid providing a magnetic field of $1.4$~$\Tesla$ aligned 
along the beam axis.
The tracking efficiency is nearly 100\% in the 
``central'' region ($\abseta<1$) and decreases towards higher
$\eta$.
Electromagnetic and hadronic calorimeters~\cite{ccal,pem} surrounding the 
solenoid measure the energies of particles up to $\abseta<3.6$. Wire chambers 
and scintillators are installed around  the hadronic calorimeter 
to detect muons with $\abseta<1.4$. Gas Cherenkov
 counters ~\cite{CLC} measure the 
average number of $\ppbar$ 
inelastic collisions per bunch crossing and thereby determine the beam luminosity.

We now outline the seven individual analyses that are then combined to achieve maximum sensitivity.
The like-sign (LS) analyses
($e^\pm e^\pm$, $e^\pm \mu^\pm$, $\mu^\pm \mu^\pm$)~\cite{lsprl}
require two leptons
of the same electric charge and do not require the detection of a third lepton.
In the trilepton analyses, the third lepton candidate can either be 
a fully reconstructed electron or muon 
($e\ell\ell$~\cite{hiptprd}, $\mu\ell\ell$~\cite{hiptprd}, $\mu\mu\ell$~\cite{loptprd}, $\ell=e,\mu$) 
or a ``track'' from a charged particle~($eet$).
The $eet$ analysis gains acceptance in detector
areas where the electron/muon detection is less efficient, and
adds sensitivity to hadronic decays of $\tau$-leptons. 
The LS, $e\ell\ell$ and $\mu\ell\ell$ 
analyses trigger on a single high $p_T$ lepton while the $\mu\mu\ell$ and
$eet$ analyses use dilepton triggers with lower lepton $p_T$ thresholds.
There is some overlap between the 
individual analyses which is taken into account for the combination: it is 
up to 30\% between some of the analyses as explained later in this letter.

The dominant SM background sources are diboson production with three or more prompt 
leptons 
($WZ/\gamma^*$, $ZZ/\gamma^*$) and Drell-Yan (DY) events 
in which the third lepton results either from the conversion of a 
bremsstrahlung photon ($Z/\gamma^*+\gamma$, 
with $\gamma\rightarrow e^+e^-$) or from a misidentified hadron.
With $Z/\gamma^*$ we denote the production of a $Z$ or a virtual photon.
For the LS analyses $W\gamma$ production with a photon conversion is also a significant background. 
Smaller background contributions
arise from heavy flavor production ($\ttbar$ and $b\bar{b}$)
with semileptonic $b$- and $c$-hadron decays.
All these background sources are modeled using several Monte Carlo (MC) event
generators. Backgrounds from $\ttbar$,  $ZZ$ and 
DY production are 
generated using {\sc pythia}~\cite{pythia}.
The $W\gamma$ and $WZ/\gamma^*$ backgrounds are generated using for the 
hard-scattering process a program by U. Baur and E.~L.~Berger~\cite{baur} and 
{\sc madevent}~\cite{madevent}, respectively. 
In all cases {\sc pythia} is used for the parton 
showering and hadronization and the parton distribution functions are 
parameterized using CTEQ5~\cite{cteq5l}. All MC events are subsequently passed through
the CDF II detector simulation based on the {\sc geant3}~\cite{geant} 
framework. 
The simulated events are then reconstructed and analyzed in 
the same way as the data.
The $b\bar{b}$ background was determined using
a combination of MC and data estimates~\cite{loptprd} 
and is negligible in most of the analyses.

An additional background source is hadrons that are 
misidentified as a lepton (``mis-id''). We determine
the misidentification 
probability in jet data samples as a function of the 
lepton $\pt$ or $\et$ using jets and tracks
and applied to the two leptons (one lepton) data 
sample for the trileptons (LS) analyses~\cite{lsprl,hiptprd}.

In the LS, $e\ell\ell$ and $\mu\ell\ell$ analyses, events are triggered on
one well-identified central electron with 
$\et>18$~GeV or muon with $\pt>18$~\GeVC.
In the $eet$ ($\mu\mu\ell$) analysis events
are triggered on two central electrons (muons)
with $\et>4$~GeV ($\pt>4$~\GeVC).  We select 
additional electrons in the central and the forward calorimeters. 
Electrons are required to have a shower shape
consistent with that expected for an electron and to have a track matched
to the calorimeter cluster. Muons
are required to deposit an amount of energy in the calorimeter
consistent with the 
expectation for a minimum ionizing particle; additionally, 
trigger muons must have associated hits in the muon detectors. 
Dedicated algorithms reject photon conversions 
and cosmic rays~\cite{hiptprd}. 
We require all leptons to be isolated from 
other particles in the event. 
For the electrons and muons in the $e\ell\ell$, $\mu\ell\ell$, $eet$ and 
$\mu\mu\ell$ analyses the isolation requirement is based only
on the calorimeter energy deposits.
For the track of the $eet$ analysis it is based only on charged tracks, and for
electrons and muons in the LS analyses it is based on both. 




We exclude events in which two leptons 
form an invariant mass $m_{\ell\ell}<15$~GeV/$c^2$ or $76<m_{\ell\ell}<106$~GeV/$c^2$
in order to remove DY and diboson events. The lower mass threshold
 additionally removes $b\bar{b}$ background.
For the $e\ell\ell$ (LS) analysis the lower mass 
value is raised to $20$ ($25$)~GeV/$c^2$ to improve the sensitivity.
For the LS analysis the mass interval near the $Z$ resonance is changed 
to $66<m_{\ell\ell}<116$~GeV/$c^2$.
Backgrounds from DY production are further reduced by requiring 
$\met>15$~GeV (for the $eet$ analysis the requirement is $\met>20$~GeV). 
The $\ttbar$ background is reduced by vetoing events 
with large hadronic jet activity.
Requirements on the angles between the leptons and $\met$ are placed
in order to reduce the cosmic-ray background.
A detailed description of the selection requirements
is given elsewhere~\cite{lsprl,hiptprd,loptprd}.

To illustrate the model sensitivity
of the search we use three example SUSY models.
The first model is the scenario of minimal supergravity (mSUGRA)~\cite{msugra},
a grand unified theory including gravity which has
five independent parameters,
fully determining all the masses and couplings of the
SUSY particles. Since the present analysis is most sensitive
to the common gaugino mass $m_{1/2}$, we fix the other four 
parameters: the common scalar mass is set to $m_0=60$~GeV/$c^2$,
the higgsino mixing parameter ($\mu$) is taken to be positive, 
the trilinear coupling ($A_0$) is set to $0$, and 
the ratio of the vacuum expectation values of the two Higgs fields 
($\tan\beta$) is set to $3$. In mSUGRA the lightest slepton is a SUSY partner of the $\tau$ 
leading to a larger branching ratio to $\tau$-leptons: about 90\% of the events
contain at least one $\tau$-lepton.
The second model we call ``MSSM ($W/Z$ model)'': all the parameters 
are taken to be the same as in the above model, but the branching ratio of $\cone(\ntwo)$ 
into leptons is fixed to be the same as the branching ratio
of the $W(Z)$ gauge boson into leptons. In the third model, the ``MSSM (no $\tilde{\ell}$-mixing)'', the
slepton chirality eigenstates are the mass eigenstates,
resulting in nearly equal branching ratios of $\cone\ntwo$ to all three lepton flavors. In this 
model  about 30\% of the events contain no $\tau$-leptons.
For all three scenarios the relationship between the masses of the gauginos is 
$m_{\cone} \approx m_{\ntwo} \approx 2 m_{\none}$. 
The slepton mass value is approximately $0.31 \cdot m_{\cone}+ 67$~GeV/c$^2$.

For the first and second models, the most stringent constraint on the chargino mass
to date is the LEP limit of $103.5$~GeV/$c^2$  
while for the third model the most stringent limit is $117$~GeV/$c^2$ as obtained by the D0 
collaboration~\cite{d0trilep}.

For the signal simulation we use {\sc softsusy}~\cite{softsusy} and {\sc isajet}~\cite{isajet} to 
compute the sparticle mass spectrum; {\sc pythia} is used to generate 
the hard-scattering events, the parton radiation, and the hadronization. CTEQ5 is used for the parton distribution functions. 
The CDF II detector's response to these events is then simulated.

%
%
The signal acceptance for the process 
$\ppbar \to \cone \ntwo+X  \to \ell\ell\ell\nu\none \none +X$, ($\ell=e,\mu,\tau$), 
in the MSSM (no $\tilde{\ell}$-mixing) scenario varies from 2.7\% 
at $m_{\cone}\approx 105$~\GeVCSq, where only the 
LS analysis is sensitive, to 6.2\% at $m_{\cone}\approx 160$~\GeVCSq,
where the acceptance of the trilepton analyses is maximal.
The acceptance in the MSSM ($W/Z$ model) 
rises from 3.0\% at low  $m_{\cone}$ to 4.0\% at 
$m_{\cone}\approx 150$~\GeVCSq.
The acceptance for mSUGRA is only 1.0\%, independent of $m_{\cone}$,
due to the enhanced decay of $\cone\ntwo$ into $\tau$-leptons.
For all models, the acceptance of the LS analyses is nearly independent of 
the chargino mass while the acceptance of the trilepton analyses
improves with increasing chargino mass.

There are systematic uncertainties on both the signal acceptance and
the background prediction that have been evaluated by many comparisons
of data and simulation~\cite{lsprl,hiptprd,loptprd}. 
The signal acceptance has an uncertainty due to the lepton 
selection~($1.5\%-13\%$) and trigger 
($<0.5\%$) efficiencies, the modeling of 
QCD radiation~($2\%-12\%$), the parton 
distribution functions~($1\%$), the integrated 
luminosity~($6\%$), and the jet energy measurement ($1\%-5\%$). The background
is also affected by the lepton- and jet-related uncertainties and the luminosity
uncertainty. Additionally, we consider uncertainties on
the lepton misidentification rate ($50\%$ for the trilepton and
$10\%-20\%$ for the LS analysis), 
the conversion background ($3\%-16\%$) and the theoretical cross sections 
($7\%$ for diboson, 10\% for $\ttbar$ and $5\%$ for DY production).
Finally, the statistical uncertainties on the Monte Carlo samples
are taken into account. 



\begin{figure}[htbp]
\begin{center}
\includegraphics*[width=0.5\textwidth]{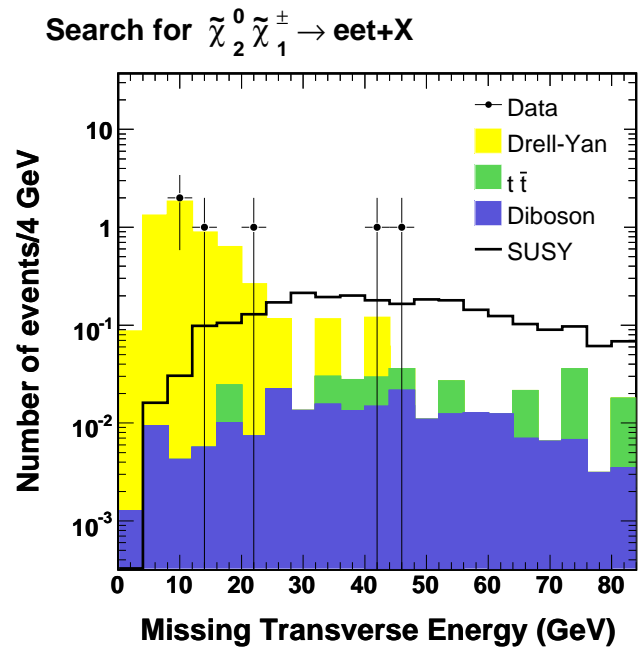}

  \caption{Missing transverse energy distribution for the $eet$ analysis
after all the selection criteria are applied except for the cut $\met>20$~GeV.
The observed data (points) are 
compared to the sum of the SM contributions (stacked, filled histograms). 
The open histogram shows the expected SUSY contribution for 
the mSUGRA scenario described in the text with $m_{\cone}=113$~\GeVCSq\ and 
$\sigma\times{\cal B} = 0.6$~pb. 
\label{fig:met}
}
\end{center}
\end{figure}

In Table~\ref{tab:bg}, the number of observed events is 
compared to the background 
contributions for each analysis.
The number of data 
events is consistent with the background expectation in all analyses
and no evidence of non-SM physics is observed. 
There is a slight excess of the data in the $e^\pm\mu^\pm$ and $eet$ analyses.
Figure~\ref{fig:met} shows the $\met$ distribution for  
the $eet$ analysis.
The observed data agree well with SM predictions at low values of $\met$ where 
the background is dominant. 
\begin{table}[htbp]
\caption{
The numbers of expected and observed events for each analysis 
before events which are shared by more than
one analysis are assigned to a single analysis.
For the $eet$ analysis the background from misidentified jets 
is included in the $Z/\gamma^*+\gamma$ background estimate of this
analysis.\label{tab:bg}.}
\vspace{1mm}
\centering{
\begin{tabular}{lrrrrrrr} \hline
 & $e^\pm e^\pm$ 
 & $e^\pm \mu^\pm$ 
 & $\mu^\pm \mu^\pm$ 
 & $e\ell\ell$  
 & $\mu\ell\ell$  
 & $eet$  
 & $\mu\mu\ell$ \\\hline
$Z/\gamma^*+\gamma$ & 0.49       & 0.62        & --     & 0.18         & 0.30  & 0.54   & 0.06      \\
$W\gamma$           & 1.54       & 1.63        & --     & --           & --    & --     & --        \\
$\ttbar+b\bar{b}$   & 0.01       & 0.03        & 0.01   & 0.04         & 0.03  & 0.22   & 0.06     \\
$WW,WZ,ZZ$          & 0.32       & 0.82        & 0.53   & 0.39         & 0.66  & 0.21   & 0.09     \\
mis-id.             & 0.60       & 0.90        & 0.38   & 0.14         & 0.27  & --     & 0.20     \\\hline
total               & 2.96       & 4.00        & 0.92   & 0.75         & 1.26  & 0.97   & 0.41     \\\hline 

uncertainty         & $\pm 0.48$ & $\pm 0.57$  & $\pm 0.12$ & $\pm 0.36$ & $\pm 0.27$  &  $\pm 0.28$  &  $\pm 0.11$ \\\hline\hline

observed                &  4         & 8           & 1       & 0            & 1     & 3      &   1        \\\hline
\end{tabular}
}
\end{table}

The data can be used to 
constrain the cross section times branching ratio, $\sigma\times\cal B$, 
and the allowed mass range of charginos for the 
SUSY scenarios discussed earlier. 

The first step in combining the results of the
seven analyses is to remove overlaps: events which are selected by more than
one analysis are assigned to the channel with the highest {\it a priori} 
signal-to-background ratio to obtain the best sensitivity.
The frequentist-based ``CL$_s$'' method~\cite{junk,read}
is then used on the resulting non-overlapping analyses.  
The correlations between the 
systematic uncertainties of the individual analyses have been evaluated 
for both the signal and the background and are taken into account
in the limit calculation.

A mass limit for the chargino is derived from the cross section limit by
comparing the observed limit
to the next-to-leading order calculation for the cross 
section~\cite{prospino}.
An uncertainty of 10\% on the signal theoretical 
cross section is included in the limit calculation~\cite{plehn}.

\begin{figure}[htbp]
  \begin{center}
    \includegraphics[width=0.5\textwidth]{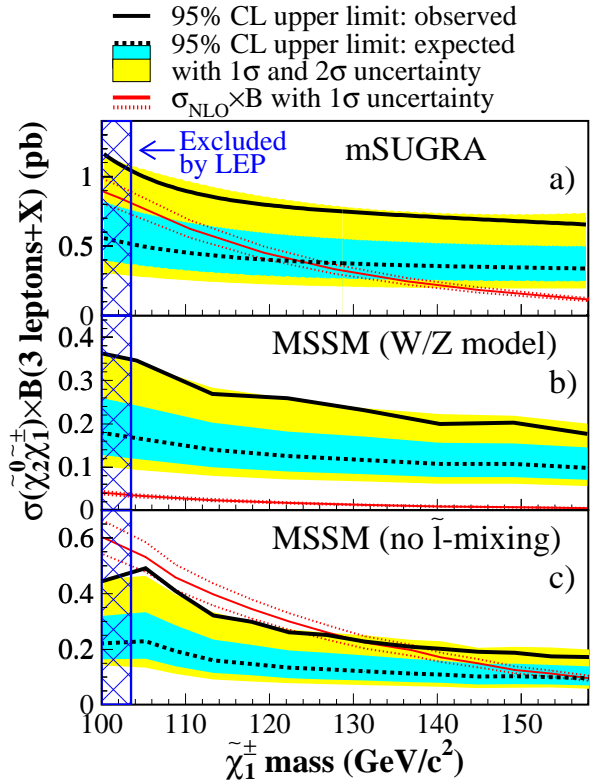}
    \caption{Cross section times branching ratio for $\ppbar \to \cone \ntwo +X \to
\ell\ell\ell+X$ production as function 
of the chargino mass. Shown are the observed and expected  
experimental upper limits together with the theoretical cross section for 
the mSUGRA (a), the MSSM ($W/Z$ model) (b), and 
the MSSM (no $\tilde{\ell}$-mixing) (c)
scenario
as described in the text. 
Also shown are the $1\sigma$ and $2\sigma$ ranges for the expected limit.\label{fig:limit} }
  \end{center}
\end{figure}

Figure~\ref{fig:limit} shows the 95\% CL upper limit 
on the  $\sigma\times {\cal B}$
as a function of the chargino mass in the three
theoretical scenarios mSUGRA, MSSM ($W/Z$ model),
and MSSM (no $\tilde{\ell}$-mixing).
In each scenario the observed and the expected limits are compared
with the theoretical $\sigma\times{\cal B}$ predictions. The expected limit is defined
to be the median limit one would obtain in a sample of 
independent experiments in which no signal is present.  
It is used to estimate the {\it a priori} sensitivity
of the search  since it does not depend on the observed data.
The observed limit is typically a factor of two larger than 
the expected, due to the fact 
that the number of events observed is higher than the predicted 
background in the $eet$ and $e^\pm\mu^\pm$ analyses.

In the mSUGRA scenario (Fig.~\ref{fig:limit}a)  
$\sigma\times {\cal B}>1$~pb is excluded for all $m_{\cone}$.
The limit improves with increasing mass down to $0.8$~pb at
$m_{\cone}=150$~\GeVCSq. 
In the MSSM $W/Z$ model and no $\tilde{\ell}$-mixing scenarios (Fig.~\ref{fig:limit}b and c)
the limits on $\sigma\times {\cal B}$ range between 0.2 and 0.4~pb. 
In the MSSM (no $\tilde{\ell}$-mixing) scenario (Fig.~\ref{fig:limit}c)
we set a 95\% CL lower limit on the chargino mass of 129~\GeVCSq, which is 
the most stringent limit to date.
In the mSUGRA and MSSM (no $\tilde{\ell}$-mixing) scenarios the expected
chargino mass limits are $122$~GeV/$c^2$ and $157$~GeV/$c^2$, respectively. 

In conclusion, we present a search for the associated production of charginos
and neutralinos using events with isolated leptons and large $\met$.
The observed data are consistent with standard model predictions. Limits
are derived on the cross section times the leptonic branching ratio
for this process and on the mass of the lightest chargino under 
different theoretical assumptions. In the MSSM (no $\tilde{\ell}$-mixing) scenario, chargino 
masses below $129$~GeV/$c^2$ are excluded at the 95\% confidence level,
extending the previous limits by $12$~GeV/$c^2$.


We thank the Fermilab staff and the technical staffs of the
participating institutions for their vital contributions. 
We thank Tilman Plehn and Benjamin Allanach for their help
on the theoretical interpretation.
This work
was supported by the U.S. Department of Energy and National Science
Foundation; the Italian Istituto Nazionale di Fisica Nucleare; the
Ministry of Education, Culture, Sports, Science and Technology of
Japan; the Natural Sciences and Engineering Research Council of
Canada; the National Science Council of the Republic of China; the
Swiss National Science Foundation; the A.P. Sloan Foundation; the
Bundesministerium f\"ur Bildung und Forschung, Germany; the Korean
Science and Engineering Foundation and the Korean Research Foundation;
the Particle Physics and Astronomy Research Council and the Royal
Society, UK; the Institut National de Physique Nucleaire et Physique
des Particules/CNRS; the Russian Foundation for Basic Research; the
Comisi\'on Interministerial de Ciencia y Tecnolog\'{\i}a, Spain; the
European Community's Human Potential Programme under contract
HPRN-CT-2002-00292; and the Academy of Finland.



\begin{thebibliography}{99}
\bibitem{cdfcoord} We use a cylindrical coordinate system where the $z$ axis is along the 
proton beam direction and $\theta$ is the polar angle.
Pseudorapidity is $\eta= -\ln \tan(\theta/2)$, while transverse momentum is 
$\pt=|p|\sin\theta$ and transverse energy is $\et= E\sin\theta$.
Missing transverse energy, $\met$, is defined as the magnitude of $-\Sigma_{i}E^{i}_{\rm T}\hat{n}_i$,
where $\hat{n}_i$ is the unit vector in the azimuthal plane that points from the beam line
to the $i^{th}$ calorimeter tower. The quantity $\et^{\mathrm{iso}}$ $(\pt^{\mathrm{iso}})$ is defined as 
the scalar sum of the transverse energies (momenta)
within a cone of
$\Delta R=\sqrt{\Delta\eta^2+\Delta\phi^2}<0.4$ around the selected 
electron (muon or track).
For electrons (muons, tracks) we require the $\et^{\mathrm{iso}}$ to be less 
than 2~GeV or 10\% of the electron $\et$ (muon or track $\pt$).
\end{thebibliography}
\end{document}